\documentclass{ws-procs975x65}

\begin{document}

\title{Generalizing the Planck distribution}

\author{Andre M. C. Souza}
\address{Departamento de Fisica, Universidade Federal de Sergipe \\
 49100-000, Sao Cristovao-SE, Brazil\\
E-mail: amcsouza@ufs.br\\
}

\author{Constantino Tsallis}
\address{Santa Fe Institute, 1399 Hyde Park Road, Santa Fe, NM, 87501 USA \\
E-mail: tsallis@santafe.edu\\
and\\
Centro Brasileiro de Pesquisas Fisicas\\ 
Rua Xavier Sigaud 150, 22290-180 
Rio de Janeiro-RJ, Brazil}

\maketitle

\abstracts{Along the lines of nonextensive
statistical mechanics, based on the entropy $S_q = k(1- \sum_i p _i^q)/(q-1) \;(S_1=-k \sum_i p_i \ln p_i)$, and Beck-Cohen superstatistics, we heuristically generalize Planck's statistical law for the black-body radiation.  The procedure is based on the discussion of the differential equation $dy/dx=-a _{1}y-(a _{q}-a_{1})\,y^{q}$ (with $y(0)=1$), whose $q=2$ particular case  leads to the celebrated law, as originally shown by Planck himself in his October 1900 paper. Although the present generalization is mathematically simple and elegant, we have unfortunately no physical application of it at the present moment. It opens nevertheless the door to a type of approach that might be of some interest in more complex, possibly out-of-equilibrium, phenomena.}


We normally obtain the statistical mechanical equilibrium distribution by
optimizing, under appropriate constraints, an entropic functional, namely the Boltzmann-Gibbs (BG) entropy $S_{BG}=-k \sum_i p_i \ln p_i$. The success and elegance 
of this variational method are unquestioned. But at least one more possibility exists, namely through differential equations. Such a path is virtually never followed. Indeed, such an approach might seem quite bizarre at first sight. But we should by no means overlook that it has at least one distinguished predecessor: Planck' s law for the black-body radiation. Indeed, Planck published two papers on the subject in 1900. The first one in October, the second one in December \cite{boya}. The bases of both of them were considered at the time as totally heuristic ones, although kind of different in nature. The second paper might be considered as a primitive form of what has now become the standard approach to statistical mechanics, based on the optimization of an entropy functional, the connection with Bose-Einstein statistics, and, ultimately, with the Boltzmann-Gibbs thermal theory for a quantum harmonic oscillator \cite{balian}. The first paper \cite{planck}, however, is totally based on simple arguments regarding an ordinary differential equation. It is along this line that the present paper is constructed. 

If $S_{BG}$ is extremized under appropriate constraints, we obtain the famous BG weight $p(E)=p(0)\,e^{-\beta E}$. This distribution can be seen as the solution of the differential equation $dp/dE=-\beta  p$ . Since more than one decade, a lot of effort is being dedicated to the study of the so called ``nonextensive  statistical mechanics", based on the generalized entropy $S_q = k(1- \sum_i p _i^q)/(q-1) \;(S_1=S_{BG})$ \cite{tsallis} (for a review, see \cite{gellmann}). The extremization of this entropy under appropriate constraints yields $p(E)=p(0)\,e_q^{-\beta E}$, where $e_q^x \equiv [1+(1-q)x]^{1/(1-q)}$ ($e_1^x=e^x$). This distribution, which has been shown to emerge in many natural and artificial systems \cite{gellmann}, can be seen as the solution of the differential equation  $d[p/p(0)]/dE=-\beta [p/p(0)]^{q}$. As a next step, we may consider even more complex systems, namely those which exhibit, for increasing $E$, a crossover from nonextensive to BG statistics. Such appears to be the case of cosmic rays \cite{cosmic}. Such situations can be handled with a differential equation which {\it unifies} the previous two ones, as follows: 
\begin{equation}
\frac{d[p/p(0)]}{dE}=-\beta _{1}[p/p(0)]-(\beta _{q}-\beta _{1})[p/p(0)]^{q} \,.
\end{equation}
Excepting for the fact that here $q$ may be noninteger, this differential equation is a particular case of Bernoulli' s differential equation. Its solution is given by 
\begin{equation}
p(E)=\frac{p(0)}{\left[ 1+\frac{\beta _{q}}{\beta _{1}}\left( e^{(q-1)\beta
_{1}E} -1\right) \right] ^{\frac{1}{q-1}}}\,,
\end{equation}
which precisely exhibits the desired crossover for $q>1$ and $0<\beta_1 <<\beta_q$ . Indeed, for $(q-1) \beta_1 E <<1$ we have that $p/p(0) \sim e_q^{-\beta_q E}$, whereas, for   $(q-1) \beta_1 E >>1$, we have $p \propto e^{-\beta_1 E}$.

In the limit $\beta_q/\beta_1 \to \infty$ and $p(0)\beta_1/\beta_q \to C$, where $C$ is a constant, Eq. (2) becomes 
\begin{equation}
p(E)=\frac{C}{\left[ e^{(q-1)\beta
_{1}E} -1 \right] ^{\frac{1}{q-1}}}\,,
\end{equation}
which, for $q=2$, becomes
\begin{equation}
p(E)=\frac{C}{ e^{\beta
_{1}E} -1  }\,.
\end{equation}
If we multiply this statistical weight by the photon density of states $g(E) \propto E^2$ and by the energy $E$, we have the celebrated frequency spectral density
\begin{equation}
u(\nu)\propto \frac{\nu^3}{e^{h \nu /k_BT}-1} \,,
\end{equation}
where we have identified $\beta_1 \to 1/k_BT$ and $E \to h\nu$. It is in this precise sense that Eq. (3) (hence Eq. (2)) can be seen as a generalization of Planck statistics. 

For $q>1$, Eq. (3) can be written as
\begin{equation}
\frac{p(E)}{C} = \sum_{n=0}^\infty d(n,q)\, e^{-\beta_1E_n} \,,
 \end{equation}
where
\begin{equation}
E_n \equiv [(q-1)n+1]\,E \propto n+\frac{1}{q-1}\,,
\end{equation}
and
\begin{equation}
d(n,q) \equiv \frac{\Gamma \left( n+\frac 1{q-1}\right) }{\Gamma \left( \frac
1{q-1}\right) \Gamma \left( n+1\right) }  \,,
\end{equation}
$\Gamma (x)$ being the Gamma function. We may now follow Planck' s path in his December 1900 paper, where he introduced the
discretization of energy that eventually led to the formulation of quantum mechanics.
Consistently, we may interpret $E_n$ as a {\it discretized energy} and $d(n,q)$ as its
{\it degeneracy}. We see that, $\forall q>1$, the spectrum is made of {\it equidistant} levels,
{\it like that of the quantum one-dimensional harmonic oscillator}. The situation is
definitively different in what concerns the degeneracy (see Fig. 1). Only for $q=2$
we have the remarkable property $ d(n,2)=1$ ($\forall n$), which recovers the harmonic
oscillator problem.

\begin{figure}[t]
\vspace{-2.5cm}
\epsfxsize=25pc \epsfbox{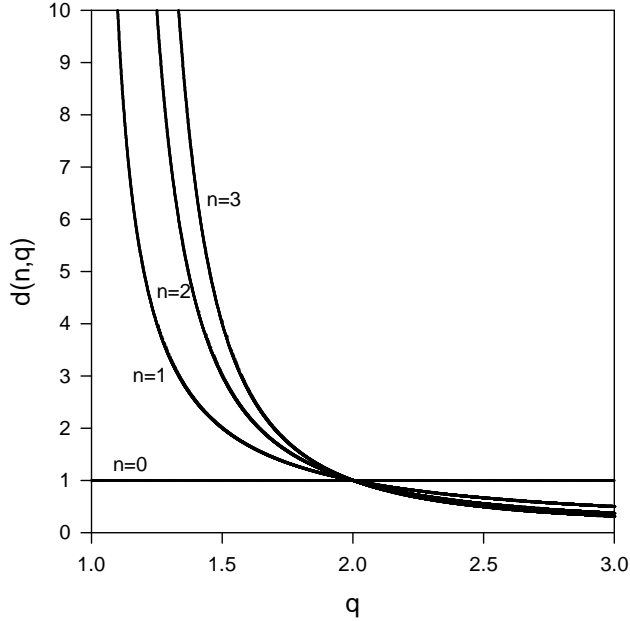}
\vspace{-3cm}
\caption{Degeneracy $d(n,q)$ as function of $q$ ($n=0,1,2,3$); $q=2$ corresponds to Planck law.}
\label{fig1.ps}
\end{figure}

At this point, let us emphasize that any thermostatistical weight (that of thermal equilibrium for instance) reflects the microscopic dynamics of the system. This fact was addressed by Einstein in 1910  \cite{einstein}, and was recently revisited by several authors (see \cite{cohen}, for instance). It was shown also, on quite general grounds, in \cite{carati}. In the same vein, a dynamical theory of
weakly coupled harmonic oscillators system was recently used  for deducing the functional relation between energy variance and mean energy that was conjectured by Einstein in connection with Planck' s formula, thus exhibiting that it is a consequence of pure dynamics \cite{galgani}. It is within this dynamical interpretation that 
Beck and Cohen introduced their {\it superstatistics}\cite{beckcohen}.  Indeed, nonequilibrium systems might exhibit spatio-temporal fluctuations of
intensive quantities, e.g., the temperature. They assumed then that the inverse temperature $\beta$ might itself be a stochastic variable, such that the generalized
distribution of energy is expressed as 
\begin{equation}
\frac{p(E)}{p(0)}=\int_{0}^{\infty }d\beta f(\beta )e^{-\beta E}\;,
\end{equation}%
where the distribution $f(\beta)$ satisfies $\int_{0}^{\infty}d\beta f(\beta )=1.$ The effective statistical mechanics of such systems
depends on the statistical properties of the fluctuations of the
temperature and similar intensive quantities. Naturally, if there are
no fluctuations of intensive quantities at all, the system must obey BG
distribution (i.e., $f(\beta )=\delta (\beta -1/k_{B}T)$). They also showed
that, if $f(\beta )$ is the $\gamma$-distribution (see also \cite{wilk}),  one obtains the $q$-exponential weight of nonextensive statistical mechanics. Moreover, for small variance of the fluctuations, the nonextensive statistical distribution  is once again 
reobtained. See  \cite{tsallisouza} for an entropic functional which, extremized under appropriate constraints, recovers the distribution of superstatistics. 

\begin{figure}
\begin{center}
\vspace{-1cm}
\includegraphics[scale=0.35]{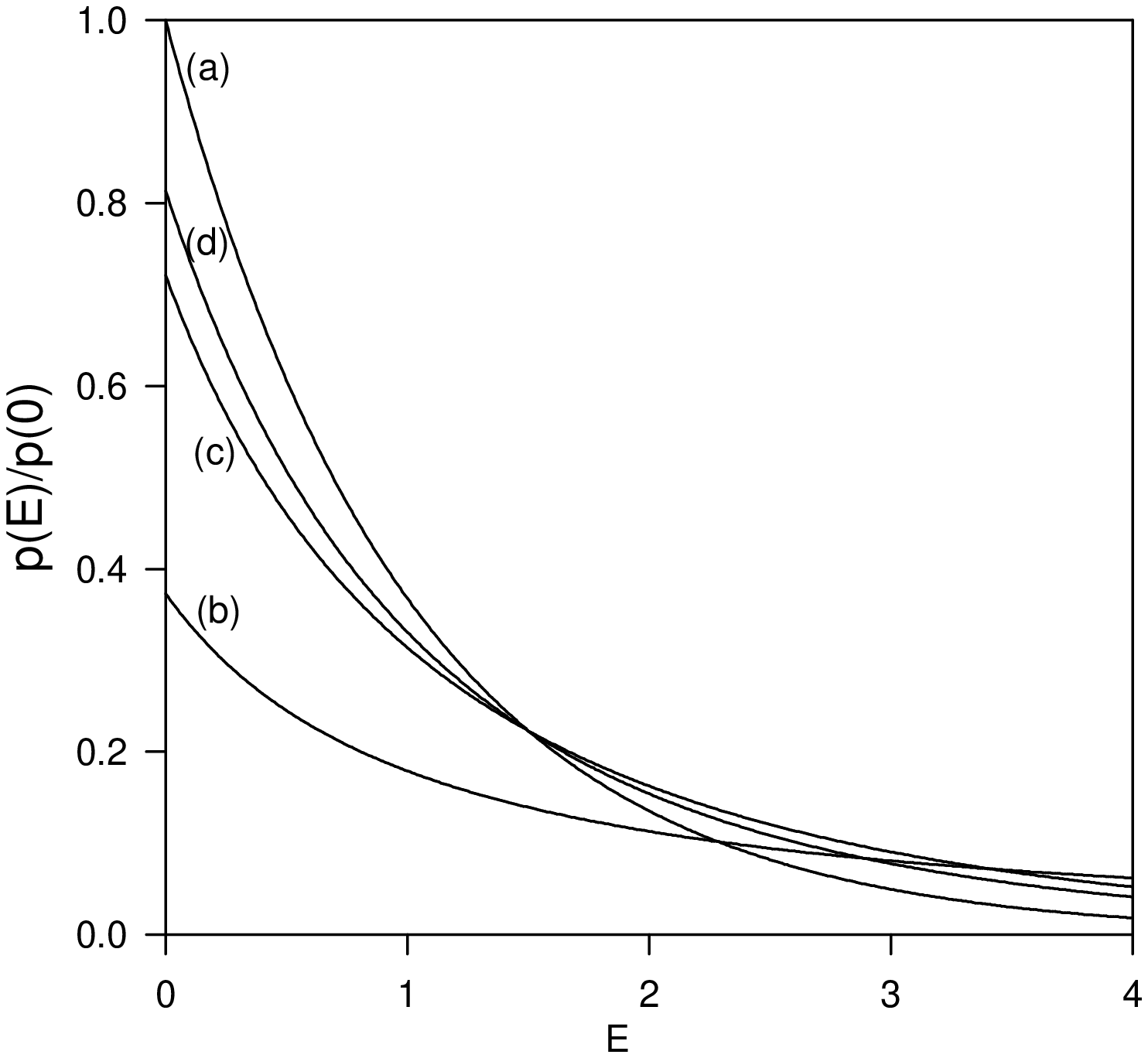}
\includegraphics[scale=0.3]{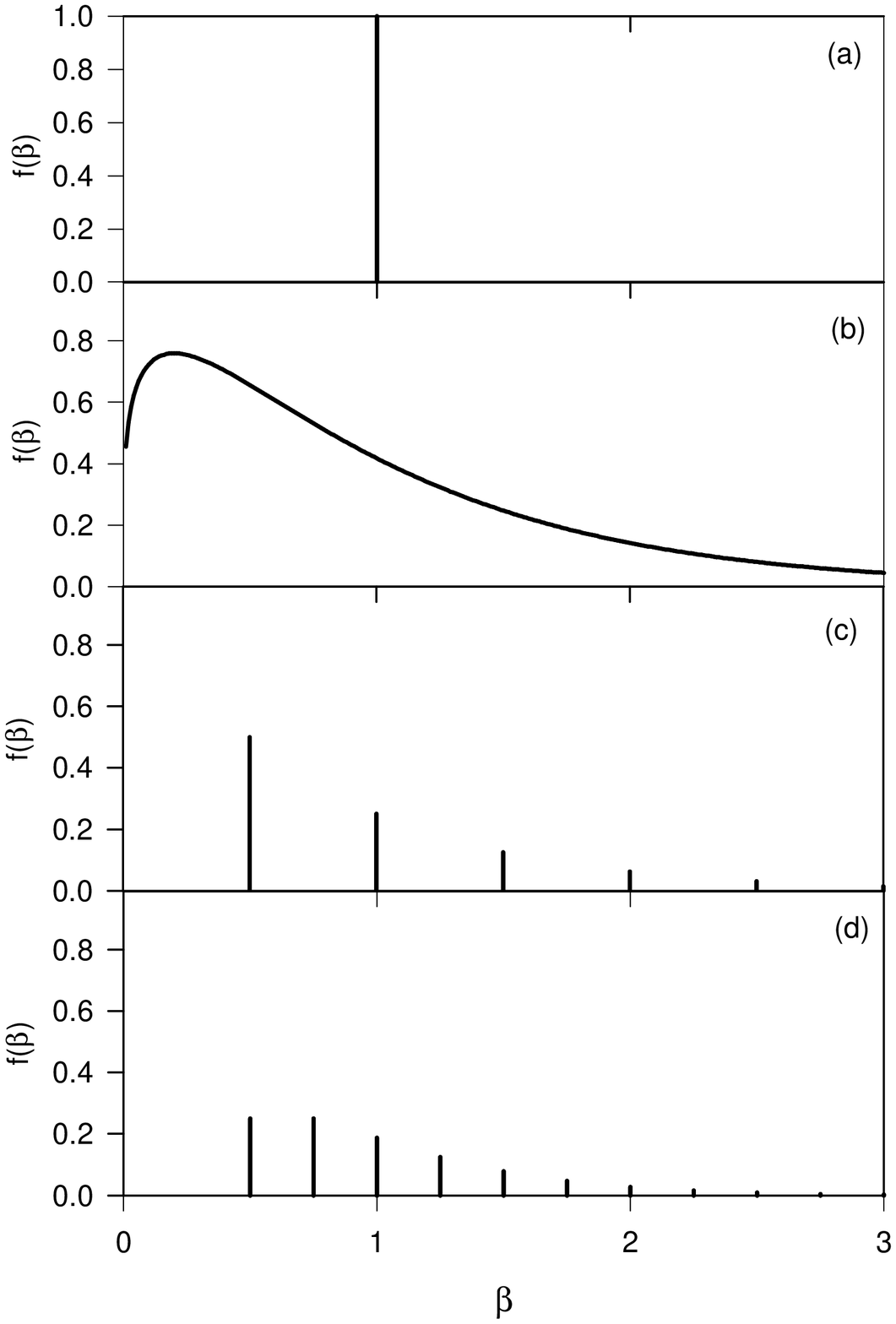}
\vspace{-2cm}
\end{center}
\caption{ Functions $\frac{p(E)}{p(0)}$ (\textit{left}) and $f(\protect\beta %
)$ (\textit{right}) for $\langle \protect\beta \rangle =1$. (a)
Boltzmann-Gibbs distribution $\Bigl[\frac{p(E)}{p(0)}=e^{-E}$; $f(\protect%
\beta )=\protect\delta (\protect\beta -1)\Bigr]$; (b) $q=q_{BC}=1.8$
distribution $\Bigl[\frac{p(E)}{p(0)}=\protect\sqrt{\frac{0.8}{\protect\pi }}%
\frac{\Gamma (1.25)}{\Gamma (0.75)}\frac{1}{(1+0.8E)^{1.25}}$; $f(\protect%
\beta )=\frac{\protect\beta ^{0.25}}{0.8^{1.25}\Gamma (1.25)}e^{-1.25\protect%
\beta }\Bigr]$; (c) $(q,q_{BC})=(2,3/2)$ distribution $\Bigl[\frac{p(E)}{p(0)%
}=\frac{(2\ln 2)^{-1}}{2e^{E/2}-1}$; $f(\protect\beta )=\frac{1}{2}[\protect%
\delta (\protect\beta -\frac{1}{2})+\frac{1}{2}\protect\delta (\protect\beta %
-1)+\frac{1}{4}\protect\delta (\protect\beta -\frac{3}{2})+...]\Bigr]$; (d) $%
(q,q_{BC})=(3/2,5/4)$ distribution $\Bigl[\frac{p(E)}{p(0)}=\frac{[4(1-\ln
2)]^{-1}}{(2e^{E/4}-1)^{2}}$; $f(\protect\beta )=\frac{1}{4}[\protect\delta (%
\protect\beta -\frac{1}{2})+\protect\delta (\protect\beta -\frac{3}{4})+%
\frac{3}{4}\protect\delta (\protect\beta -1)+...]\Bigr]$. In the cases
(a,c,d), what is represented is not $f(\protect\beta )$ strictly speaking,
but rather the weights of the Dirac delta's. }
\end{figure} 
We straightforwardly obtain, through Laplace transform, that the superstatistical distribution $f(\beta )$
corresponding to the $p(E)/p(0)$ given by Eq. (2) is 
\begin{equation}
f(\beta )=\left( \frac{\beta _{1}}{\beta _{q}}\right) ^{\frac{1}{q-1}%
}\sum\limits_{n=0}^{\infty }d(n,q)\left( 1-\frac{\beta _{1}}{\beta _{q}}%
\right) ^{n}\delta \left( \beta -\beta _{1}[(q-1)n+1]\right) .
\end{equation}
Moreover, we define  
\begin{equation}
q_{_{_{BC}}} \equiv \frac{\langle \beta ^{2}\rangle }{\langle \beta \rangle ^{2}} \,,
\end{equation}
where $\langle ...\rangle \equiv \int_{0}^{\infty }d\beta f(\beta )(...)$.  
The notation $q_{_{BC}}$ ({\it BC} stands for {\it Beck-Cohen}) has been introduced to avoid
confusion with the present $q$. Only when $f(\beta)$ equals the $\gamma$-distribution we
 have $q_{BC}=q$. Using Eq. (10)  and integrating 
we obtain 
\begin{equation}
\langle \beta \rangle =\beta _{q}\qquad \text{and\qquad }\langle \beta
^{2}\rangle =\beta _{q}^{2}\left( q-(q-1)\frac{\beta _{1}}{\beta _{q}}
\right) .
\end{equation}
Replacing(12) into (11) we obtain 
\begin{equation}
\frac{\beta _1}{\langle \beta \rangle}=  \frac{q-q_{_{BC}}}{q-1}\;\;\;\;(1 \le q_{BC} \le q)\, .
\end{equation}

It is worthy remarking that, for all admissible $f(\beta )$, we can write the asymptotic expression $p(E)/p(0)=\langle e^{-\beta E}\rangle \sim e^{-\langle \beta \rangle E}(1+\frac{
\sigma ^2E^2}2)$, where $\sigma \equiv \sqrt{\langle \beta ^2\rangle -\langle \beta
\rangle ^2}=(q_{_{BC}}-1)\langle \beta \rangle \to 0$. 

Finally, we may rewrite distribution (2) as follows:

\begin{equation}
\frac{p(E)}{p(0)} =\left[ 1+\frac{q-1}{q_{_{BC}}-q}\left( 1-e^{(q-q_{_{BC}})\langle \beta
\rangle E}\right) \right] ^{-\frac 1{q-1}}\text{,}
\end{equation}
hence, through Laplace transform,
\begin{equation}
f(\beta )=\left( \frac{q_{_{BC}}-q}{1-q}\right) ^{\frac{1}{q-1}%
}\sum\limits_{n=0}^{\infty }d(n,q)\left( \frac{q_{_{BC}}-1}{q-1}\right)
^{n}\delta \left[ \beta -\langle \beta \rangle \frac{(q-q_{_{BC}})}{(q-1)}%
[(q-1)n+1]\right]   \label{disc}
\end{equation}

Observe that, for all $q$, if $q_{_{BC}}\rightarrow 1$ we obtain the BG
distribution. In addition, we see that $\beta $ generically assumes discrete
values in $f(\beta )$ . If we focus on the limit of continuous values for $
\beta $, we must have (using Eq. (10)) $\Delta \beta \equiv  \beta
(n+1)-\beta (n) =\beta _{1}(q-1)\rightarrow 0$, and this is obtained
(see Eq. (13)) when $\langle \beta \rangle \to 0$ (i.e., high temperature) or $q_{_{BC}}\rightarrow q$ (i.e., $q$-statistics) .
 In Fig. 2 we present typical examples
of pairs $\bigl(p(E)/p(0),f(\beta )\bigr)$.

Summarizing, we obtained the distribution corresponding to the differential equation (1), expected to characterize a class of physical stationary states where a crossover occurs between nonextensive and BG statistics. This led us to a possible generalization of Planck law. We obtained also the Beck-Cohen superstatistical distribution $f(\beta)$ associated with such type of crossovers between statistics. Along similar lines, it is possible to study crossovers between $q$ and $q^\prime$ statistics, with eventual applications in turbulence and other complex phenomena. 

\section*{Acknowledgments}

Partial support from PCI/MCT, CNPq, PRONEX, FAPERJ and FAP-SE (Brazilian
agencies) is acknowledged.

\end{document}